\documentclass[aps,twocolumn,preprintnumbers]{revtex4}
\usepackage{epsfig}
\usepackage{amssymb,amsmath,amsfonts,amsthm,graphicx,psfrag}
\graphicspath{{./Fig/}}                 
\newcommand{\noi}{\noindent}
\newcommand{\beq}{\begin{equation}}
\newcommand{\eeq}{\end{equation}}
\newcommand{\bea}{\begin{eqnarray}}
\newcommand{\eea}{\end{eqnarray}}
\newcommand{\beas}{\begin{eqnarray*}}
\newcommand{\eeas}{\end{eqnarray*}}
\newcommand{\Fig}[1]{Fig.~\ref{#1}}
\newcommand{\Tab}[1]{Table~\ref{#1}}
\newcommand{\Sec}[1]{Section~\ref{#1}}
\newcommand{\Eq}[1]{Eq.~(\ref{#1})}

\newcommand{\aleq}{\mbox{}_{\textstyle \sim}^{\textstyle < }}
\newcommand{\ageq}{\mbox{}_{\textstyle \sim}^{\textstyle > }}

\newcommand{\Nt}{N_{\tau}}
\newcommand{\Ns}{N_\sigma}
\newcommand{\cV}{\mathcal{V}}
\newcommand{\cB}{\mathcal{B}}
\newcommand{\cE}{\mathcal{E}}

\begin{document}
\preprint{HU-EP-13/57}
\title{Magnetic catalysis (and inverse catalysis) at finite temperature 
in two-color lattice QCD} 
\author{E.-M. Ilgenfritz}
\affiliation{Joint Institute for Nuclear Research, VBLHEP, 141980 Dubna, Russia}
\author{M.~M\"uller-Preussker}
\affiliation{Humboldt-Universit\"at zu Berlin, Institut f\"ur Physik, 
12489 Berlin, Germany}
\author{B.~Petersson}
\affiliation{Humboldt-Universit\"at zu Berlin, Institut f\"ur Physik, 
12489 Berlin, Germany}
\author{A.~Schreiber}
\affiliation{Humboldt-Universit\"at zu Berlin, Institut f\"ur Physik, 
12489 Berlin, Germany}

\date{February 4, 2014}

\begin{abstract}
Two-color lattice QCD with $N_f=4$ staggered fermion degrees of freedom
(no rooting trick is applied) with equal electric charge $q$ is studied
in a homogeneous magnetic background field $B$ and at non-zero temperature 
$T$. In order to circumvent renormalization as a function of the bare coupling
we apply a fixed-scale approach. We study the influence of the magnetic
field on the critical temperature. At rather small pseudo-scalar meson 
mass ($m_{\pi} \approx 175~\mathrm{MeV} \approx T_c(B=0)$) we confirm 
a monotonic rise of the quark condensate $<\bar{\psi} \psi>$ with increasing 
magnetic field strength, i.e. {\it magnetic catalysis}, as long as one is
staying within the confinement or deconfinement phase. In the transition
region we find indications for a non-monotonic behavior of $T_c(B)$ 
at low magnetic field strength ($qB~<~0.8~\mathrm{GeV}^2$) 
and a clear rise at stronger magnetic field.   
The conjectured existence of a minimum value $T_c(B^{*}) < T_c(B=0)$ would
leave a temperature window for a decrease of $<\bar{\psi} \psi>$ with 
rising $B$ ({\it inverse magnetic catalysis}) also in the present model.        
\end{abstract}

\maketitle

\section{Introduction}
\label{sec:introduction}
The interaction of strong magnetic fields with hadronic matter
has recently been widely discussed because of its relevance to 
non central heavy ion collisions. In such collisions there will be two
lumps of spectators moving in opposite directions. They give rise to a 
magnetic field perpendicular to the reaction plane, which may be estimated
from the Lienard-Wiechert potentials of the moving spectators. 
From these estimates it can be shown that the magnetic field is
so strong that its consequences cannot be studied perturbatively. 
In fact, the field is estimated to have strength 
$eB\sim  m_{\pi}^2\sim 10^{18}$ Gauss at RHIC and LHC at the time of 
formation of the fireball. The field strength falls for large time $t$
at most as $1/t^2$ and because of the effect of electrical conductivity
may reach a plateau \cite{Kharzeev:2007jp,Skokov:2009qp,McLerran:2013hla}. 
Therefore, for a longer time reaching from the formation of the fireball
to the final transition from quark-gluon to hadron matter, it may be a
reasonable approximation to assume a constant external field excerting
influence on the transition.

It is known since a long time that the problem of a relativistic particle 
with spin $0$  or $1/2$ in a constant external magnetic field can be solved 
exactly \cite{Geheniau:1950xx,Katayama:1951xx,Schwinger:1951nm}. With the 
help of these solutions one can discuss the effect in the Nambu-Jona-Lasinio 
\cite{Klevansky:1989vi} or in the chiral \cite{Shushpanov:1997sf} model 
at zero temperature. The general result is that the magnetic field induces 
an increase of the chiral condensate. 
This was dubbed {\it magnetic catalysis} in Ref. \cite{Gusynin:1994re} 
and claimed to be essentially model independent.
For a recent review see \cite{Shovkovy:2012zn}. The model calculations have 
been extended to finite temperature, in order to study the phase diagram of 
strongly interacting matter in a constant magnetic field. In this case 
there is no claim of model independence. The critical temperature of the 
chiral phase transition rises in most calculations \cite{Agasian:2001hv}. 
There are also claims that the chiral and the deconfinement phase transitions 
split, and the latter decreases with the magnetic field strength 
\cite{Agasian:2008tb}.

Recently several groups have started to investigate the problem through
ab initio lattice simulations of QCD and QCD-like theories in a constant 
external magnetic field. There is no sign problem in contrast to e.g.
the introduction of a chemical potential in QCD.
The pioneering work was performed by M. Polikarpov and collaborators
\cite{Buividovich:2008wf,Buividovich:2009ih,Buividovich:2009wi,
Buividovich:2009my}.  They carried out their calculations in quenched $SU(2)$. 
In our previous paper \cite{Ilgenfritz:2012fw} we extended the 
calculations to $SU(2)$ with four flavors of dynamical fermions. The choice 
of four flavors eliminates the need for rooting of the Kogut-Susskind or 
staggered fermion action, the latter being still under debate.
But one should notice that in this case we expect a first order finite 
temperature transition \cite{Pisarski:1983ms} in contrast to the observed 
smooth crossover in the $N_f=2$ as well as $2+1$ cases of full QCD at 
non-vanishing $u$-, $d$-quark mass. In  Ref. \cite{Ilgenfritz:2012fw} we 
reported {\it magnetic catalysis} for all temperatures investigated. 
The deconfinement transition, which we determined from the behavior of the
Polyakov loop and the various parts of the gluonic action coincided within our 
precision with the chiral transition. The transition temperature increased with 
increasing magnetic field. However, in our previous calculations 
the temperature dependence was studied only by 
varying the bare coupling parameter $\beta$, while the magnetic field strength
as well as the fermion mass was fixed in lattice units. As a consequence the
physical field strength as well as the fermion mass was increasing with the
temperature. This disadvantage is avoided in our present paper. 

Two groups have performed simulations in full QCD in the presence of a magnetic
field (\cite{D'Elia:2010nq,D'Elia:2011zu,Bonati:2013lca} and
\cite{Bali:2011qj,Bali:2012zg,Bali:2012jv,Bali:2013esa,Bruckmann:2013oba},
see also \cite{Levkova:2013qda}). 
Both groups observe {\it magnetic catalysis} for temperatures in the confined 
phase. Near the phase transition only the second group observes what they call 
{\it inverse magnetic catalysis},  i.e. the chiral condensate and thus the 
transition temperature decreases with increasing magnetic field strength 
\cite{Bali:2011qj}. It is still not completely clear, whether the discrepancy 
is explained by the different sets of quark masses used. In \cite{D'Elia:2011zu} 
the case of two flavors is treated and the parameters chosen lead to a pion 
mass of approximately 200 MeV. In \cite{Bali:2011qj} the parameters and 
the action used are the same as in \cite{Borsanyi:2010cj}, namely $2+1$ 
flavors with the parameters chosen to give the physical mass to the
Goldstone pion connected to the exact lattice axial symmetry $U(1)$. 
In both calculations, the fourth root of the fermion determinants
is taken to reduce the number of flavors (also called tastes). This procedure 
is still under debate. A nice recent review of the lattice results for QCD and 
QCD-like theories in external fields can be found in Ref. \cite{D'Elia:2012tr}.

In this article we extend our calculations in \cite{Ilgenfritz:2012fw} of the 
two-color theory with four flavor fermion degrees of freedom with equal 
electric charges to a considerably smaller value of the bare 
quark mass. In fact, now the ratio of the Goldstone pion mass to the critical 
temperature is similar to the physical case of QCD. Furthermore, we use the 
fixed-scale approach, which means that the lattice spacing dependence of the 
renormalization factors is irrelevant for our results. We measure the various 
parts of the gluon action, the Polyakov loop and the chiral condensate.
With the help of these measurements we localize the finite temperature 
transition, and describe its dependence on the magnetic field strength. 
Although our model is not QCD, the chiral properties are quite similar. 
Furthermore, investigations of the dynamical $SU(2)$ theory are of considerable 
interest, because they can be extended to finite chemical potential 
without a sign problem. It is also easier to investigate the topological 
structure of the lattice gauge fields than in the $SU(3)$ case.

In Section II, for completeness, we specify the action and the order 
parameters, although they are the same as in our previous calculation 
\cite{Ilgenfritz:2012fw}. In Section III we describe the simulation 
parameters, and in Section IV the scale determination. Section V is 
devoted to a presentation of our finite temperature results. 
Finally, in Section VI we discuss the results, 
compare with results of other groups, and present our conclusions.

\section{Specification of the action and order parameters}
\label{sec:specifications}

The theory, which we have chosen to investigate, is color $SU(2)$ with 
four fermion flavors. We want to study its behavior at finite temperature
under the influence of a strong external magnetic field. To this end we perform
numerical simulations in the lattice regularization, which are fully
non-perturbative also in the electromagnetic coupling to the magnetic field.
The details of the corresponding model on the lattice are given in
\cite{Ilgenfritz:2012fw}. For completeness we present again the main building 
blocks here.

We introduce a lattice of four dimensional size
\beq
\cV\equiv \Nt\times \Ns^3.
\eeq
The sites are 
enumerated by $n=(n_1,n_2,n_3,n_4)$, where the $n_i$ are integers, 
$n_i =1,2, \ldots ,\Ns$ for $i=1,2,3$ and $n_4 = 1,2, \ldots , \Nt$.
The fourth direction is taken as the Euclidean time 
direction. The lattice spacing is denoted by $a$. The physical volume $V$ and 
the temperature $T$ of the system are given by 
\bea
V & = & (a\Ns)^3\,, \\
T & = & \frac{1}{a\Nt}.
\eea

On the links 
$n \to n+\hat{\mu}$ we define group elements $U_{\mu}(n)\in SU(2)$, 
where $\mu = 1,2,3,4$.  The boundary conditions of the $U$-fields 
are periodic. For the gauge part of the action
we choose the usual Wilson action, 
\beq
S_G= \beta\cV \sum_{\mu<\nu}P_{\mu\nu},
\eeq
where
\beq
P_{\mu\nu} = \frac{1}{\cV}\sum_n(\frac{1}{2}Tr\left(1-U_{\mu\nu}(n)\right))
\eeq
\noi
with $U_{\mu\nu}(n)$ denoting the $\mu\nu$-plaquette matrix attached to 
the site $n$.
\noi

For the fermion part of the action, we use staggered fermions, which are 
spinless Grassmann variables $\bar{\psi}(n)$ and $\psi(n)$ 
being vectors in the fundamental representation of the gauge group $SU(2)$.
The different flavor degrees of freedom are assumed to carry equal 
electric charges allowing to interact with an external magnetic field. 
The boundary conditions of the fermionic fields are periodic
in the space directions and antiperiodic in the time direction. 
In the absence of a magnetic field the fermionic part of the action 
which we use becomes the usual staggered action,
\beq
S_F  =  a^3\sum_{n,n^{\prime}} \bar{\psi}(n)[D(n,n^{\prime}) 
     + ma\delta_{n,n^{\prime}}
]\psi(n^{\prime}),
\eeq
where $ma$ is the bare quark mass and
\bea 
D(n,n^{\prime}) 
& = & \frac{1}{2}\sum_{\mu}\eta_{\mu}(n)[U_{\mu}(n)\delta_{n+\mu,n^{\prime}}-  
\nonumber \\
& - & U_{\mu}^{\dagger}(n-\mu)\delta_{n-\mu,n^{\prime}}]\,. \label{stag}
\eea

\noi
The arguments $n,~n^{\prime}$ are integer four-vectors 
denoting sites on the lattice and $\eta_{\mu}(n)$
are the normal staggered sign factors,
\bea
\eta_1(n) & = & 1\,, \nonumber \\ 
\eta_{\mu}(n) & = & (-1)^{\sum_{\nu=1}^{\mu-1}} n_{\nu}\,, \hspace{1cm} 
\mu=2,3,4\,.
\eea

We introduce electromagnetic 
potentials in the fermion action by new, commuting group elements on the
links, namely $V_{\mu}(n)= e^{i\theta_{\mu}(n)}\in U(1)$.
As discussed in our earlier work \cite{Ilgenfritz:2012fw}
a constant magnetic background field in the $z \equiv 3$-direction 
going through all the $(x,y) \equiv (1,2)$ -planes of finite size 
$\Ns \times \Ns$ with a constant magnetic flux $\phi =a^2qB$ 
through each plaquette can be realized as follows:
\bea
&&V_1(n)  =  e^{-i\phi n_2/2} \hspace{0.4cm} (n_1 =1,2,\ldots , \Ns-1)\,, 
\nonumber \\ 
&&V_2(n)  =  e^{i\phi n_1/2} \hspace{0.5cm} (n_2=1,2,\ldots ,\Ns-1)\,, 
\nonumber \\
&&V_1(\Ns,n_2,n_3,n_4)  =  e^{-i\phi(\Ns+1) n_2/2}\,,  \\
&&V_2(n_1,\Ns,n_3,n_4)  =  e^{i\phi(\Ns+1) n_1/2}\,, 
\nonumber \\
&&V_3(n)=V_4(n)=1\,. \nonumber 
\eea

\noi
With periodic boundary conditions the magnetic flux becomes quantized 
in units of $2 \pi / \Ns^2$,
\beq
\phi=a^2qB= \frac{2\pi N_b }{\Ns^2}\,, \hspace{1cm} N_b\in Z. 
\label{flux}
\eeq
Because the angle $\phi$ is periodic there is an upper bound on the
flux $\phi < \pi$. In practice, to avoid finite-size effects we restrict
ourselves to $\phi < \pi/2$. Inserting this into (\ref{flux}) one 
obtains the condition
\beq
N_b < \Ns^2/4\,.
\eeq
Thus at finite temperature, $\frac{\sqrt{qB}}{T}$ (for $qB > 0$) is 
restricted to the region 
\beq
\sqrt{2\pi} \frac{\Nt}{\Ns} \le \frac{\sqrt{qB}}{T} < \sqrt{\frac{\pi}{2}}\Nt. 
\label{eq:limits}
\eeq

Finally we introduce the fields $V_{\mu}(\theta)$ into the fermionic action
(\ref{stag}) by substituting
\bea
U_{\mu}(n) & \rightarrow & V_{\mu}(n)U_{\mu}(n)\,, \\
U_{\mu}^{\dagger}(n) & \rightarrow & V_{\mu}^{\ast}(n)U^{\dagger}_{\mu}(n)\,. 
\eea
The partition function is given by
\beq
Z(\theta)=\int \prod(d\bar{\psi}(n)d\psi(n) dU_{\mu}(n))e^{-S_G - S_F(\theta)}.
\eeq
Note that the fields $\theta_{\mu}(n)$ are not treated as dynamical variables,
and that there is thus no corresponding dynamical part of the action.

To determine the lattice spacing we calculate the potential between 
heavy quarks on a zero temperature lattice at vanishing magnetic field. 
On the same lattice we also measure the Goldstone pion mass. 
Details of these calibration measurements are given in 
\Sec{sec:scale} below.

To study the influence of an external magnetic field on two-color QCD
at finite temperature, we shall first look at the anisotropy in the gluonic
action by measuring the average value $<P_{\mu\nu}>$ of the non Abelian 
plaquette energies for the different combinations of directions. 

We further measure the following approximate order parameters.

The chiral condensate, which is an exact order parameter in the 
limit of vanishing quark mass, is given by

\bea
a^3<\bar{\psi}\psi>  =  - \frac{1}{\cV}~\frac{1}{4}
~\frac{\partial}{\partial(ma)} \log (Z) = \nonumber \\
 =  \frac{1}{\cV}~\frac{1}{4}~<\mathrm{Tr}(D+ma)^{-1}>.
\eea
The factor $1/4$ is inserted because we define $<\bar{\psi}\psi>$ per flavor,
and our theory has $4$ flavors.

We compute also the average value of the Polyakov loop $<L>$, which is 
the order parameter for confinement in the limit of infinite quark mass
(the pure gauge theory), 
\bea
 &  <L> = & \frac{1}{\Ns^3}\sum_{n_1,n_2,n_3} \frac{1}{2} \times \\ 
 & &  <\mathrm{Tr} \left( \prod_{n_4=1}^{\Nt} U_4(n_1,n_2,n_3,n_4)\right)>\,.
\nonumber
\eea
It is important to notice that the 
mean values defined above are bare quantities
which should be renormalized when comparing with continuum 
expectation values.

\section{Simulation setup}
\label{sec:setup}

In the present investigation we use the fixed-scale approach, i.e. we keep
$\beta$ fixed and thereby the lattice spacing $a$ and vary
the temperature by changing $\Nt$. More precisely we simulated the theory 
at $\beta=1.80$ mainly with lattice sizes $32^3\times \Nt\,,~\Nt=4,6,8,10$,  
and with a lowest mass value $ma=0.0025$ taking each time at least
three values of the magnetic flux, $N_b=0, 80, 200$.

The simulation algorithm employed is the usual Hybrid Monte Carlo method,
updated in various respects in order to increase efficiency
(even-odd and mass preconditioning, multiple time scales, Omelyan integrator 
and written in CUDA Fortran for the use on GPU's). The number of configurations 
(trajectories) generated in a simulation varied between 3000 and 5000. 
In general, 250 configurations were discarded for initial thermalization.

We measured the chiral condensate on every third configuration, apart from 
$\Nt=6$ and $N_b=80, 200$ where we used every fifth configuration only,
because in these runs we are close to the transition temperature. The Polyakov 
loop and the plaquette variables were measured on every configuration.
The chiral condensate was evaluated with the random source method. Thereby we
used 100 $Z_2$ random sources per configuration. The integrated autocorrelation 
times of all observables were taken into account in the error analysis. 
It could be estimated to be mostly well below 20 consecutive trajectories.

A zero temperature simulation with zero magnetic field was performed for the 
same $\beta=1.80$ and for the two mass values $ma=0.0025,~0.01$ on a lattice of 
size $32^3\times 48$ in order to estimate the lattice spacing and the
pion mass. The number of trajectories in this run was about 
750 and the first 200 were discarded. Measurements were performed after 
every third trajectory. 

\section{Fixing the lattice scale}
\label{sec:scale}

In order to determine the lattice spacing we investigate 
the potential between infinitely heavy quarks. We use the
Sommer parameter, defined in the continuum by 
the equation
 
\beq
\left. r^2\,\frac{dV}{dr}\right|_{r=r_0}=1.65 \label{r0}\,.
\eeq

\begin{figure*}[h!t]
\centering
\includegraphics[width=\textwidth]{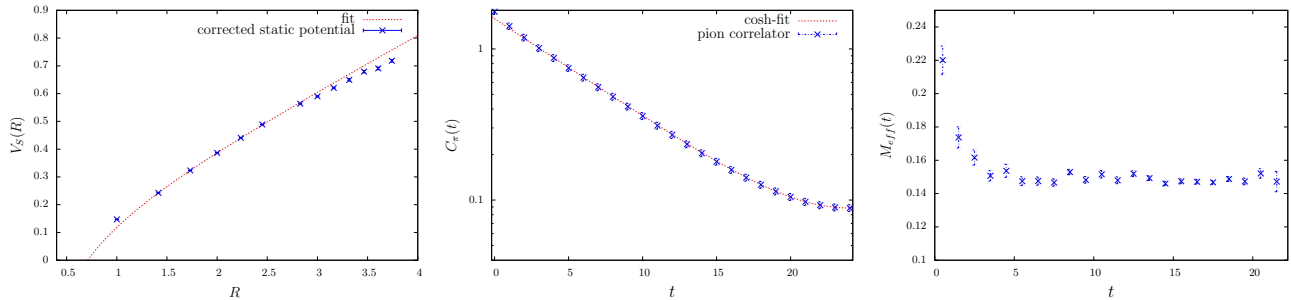}
\caption{Potential of a static quark-antiquark pair (left), pion correlator 
(middle) and effective mass $M_{eff}(t)$ (right) all measured at $\beta=1.80$ 
and bare quark mass $am=0.0025$. The lattice size is $32^{3} \times 48$.
The dotted line in the left panel corresponds to a fit the parameters of which 
are given in \Tab{tab1}.
}
\label{fig1}
\end{figure*}

On the lattice we measure the potential using Wilson loops. 
In order to increase the 
signal-to-noise ratio, HYP-smearing \cite{Hasenfratz:2001hp} with 
additional APE-smearing \cite{Albanese:1987ds} (in the version used in 
\cite{Bornyakov:2005iy}) was applied to the gauge configurations before 
measurements were performed. The potential $V(\vec{R})$ as extracted 
from Wilson loops is not spherically symmetric, in particular for small 
distances. Defining $R$ as the distance in lattice units ($r=Ra$), 
we introduce a more symmetric potential $V_S(R)$ by
\cite{Michael:1992nj,Bali:1992ru}

\beq
V(\vec{R})=V_S(R)+C(\frac{1}{R}-G_L(\vec{R}))\,, \label{eq:V}
\eeq
where $G_L(\vec{R})$ is the free gluon propagator on the lattice.
We then make the Ansatz
\beq
V_S(R)=A_1-\frac{A_2}{R} +\sigma a^2 R \label{eq:VS}\,.
\eeq
The potential $V_S(R)$ as well as the best fit are shown in 
\Fig{fig1} (left panel).
The fit parameters are given in \Tab{tab1}.
\begin{table}[h!]
\centering
\mbox{\scriptsize
\setlength{\tabcolsep}{1.0pt}
\begin{tabular}{|c|c|c|c||c|c|c|c|c|c|c|}
\hline
 $\beta$ & $a m$ & $\Ns$ & $\Nt$ & $R_1$ & $R_2$ & C
                 & $A_{1}$ & $A_{2}$ & $\sigma a^2$ & $\chi^2_{dof}$ \\
\hline
  1.8 & .01   &  16 & 32 & 1.2  & 3.2 & .167(25) & .265(38) & .370(42) &  .169(8) &  0.87  \\
  1.8 & .0025 &  32 & 48 & 1.2  & 3.0 & .083(21) & .078(24) & .152(21) &  .192(7) &  1.16  \\ 
\hline
\end{tabular}
}
\caption{Fit parameters in lattice units for the static potential $V(\vec{R})$ acc. 
to Eqs. (\ref{eq:V}) and (\ref{eq:VS}) for two sets of parameters considered 
in~\cite{Ilgenfritz:2012fw} and in this work. $R_1$ and $R_2$ define the fit 
range for the static potential in lattice units.}
\label{tab1}
\end{table}
\begin{table}[h!]
\centering
\mbox{\scriptsize
\setlength{\tabcolsep}{1.0pt}
\begin{tabular}{|c|c|c|c||c|c|c|c|}
\hline
 $\beta$ & $a m$ & $\Ns$ & $\Nt$ & $t_{min}$ &$C_{0}$ & $E=am_\pi$ &  $\chi^2_{dof}$ \\
\hline
  1.8 & .01   & 16 & 32 & 6 &1.01(3)   & 0.285(1) & 0.023   \\
  1.8 & .0025 & 32 & 48 & 7 &1.58(8)   & 0.149(3) & 0.010     \\ 
\hline
\end{tabular}
}
\caption{Fit parameters in lattice units for the pion correlator $C_{\pi}(t)$ 
according to \Eq{eq:pionfit} for the two sets of simulation parameters 
considered in~\cite{Ilgenfritz:2012fw} and in this work. 
The fit range for the pion correlator starts at lattice distance $t_{min}$.}
\label{tab2}
\end{table}
\begin{table}[h!]
\centering
\mbox{\scriptsize
 \begin{tabular}{|c|c|c|c|c||c|c|c|c|}
\hline
 $\beta$ & $a m$ & $\Ns$ & $\Nt$ & $N_b^m$ & $R_0$ 
         & $a [\mathrm{fm}]$ & $m_\pi[\mathrm{MeV}]$ & $\sqrt{qB}_{m}[\mathrm{GeV}]$ \\
 \hline
  1.8 & .01  & 16 & 32 & 50  & 2.75(8)  & 0.170(5)  & 330(10)  & 1.29(4)  \\ 
  1.8 & .0025& 32 & 48 & 200 & 2.78(6)  & 0.168(4)  & 175(4)   & 1.30(3)  \\ 
\hline
\end{tabular}
}
\caption{Results for the Sommer scale $R_0$ (in lattice units), the 
lattice spacing $a$, the pion mass $m_\pi$, and the quantity $\sqrt{qB}_m$ 
characterizing the magnetic field strength for the largest number of flux units 
$N_b^m$ used for various setups of simulation parameters considered 
in \cite{Ilgenfritz:2012fw} and this work.}
\label{tab3}
\end{table}
As the two sides in \Eq{r0} are dimensionless, the same equation holds 
for the lattice distance $R$. Thus, assuming the form (\ref{eq:VS}) for the 
potential $V$ in (\ref{r0}) we obtain 
\beq
r_0/a=R_0=\sqrt{\frac{1.65-A_2}{\sigma a^2}}\,. \label{R0}
\eeq

We are, of course, aware of the fact that we are considering a fictious world
of two-color QCD with four flavors of quarks with equal charges q. 
Nevertheless, the scale determination provides a rough estimate of the 
magnetic field strength for the various values of the flux and the distance 
to the chiral limit. 

Inserting the value $r_0=0.468(4)$ fm  \cite{Bazavov:2011nk} 
we obtain the lattice spacing from a fit with formula (\ref{R0}) 
\beq
a=0.168(4)~\mathrm{fm}  \label{fm}
\eeq
for $ma=0.0025$ and an only slightly larger value 
for $ma=0.01$ \cite{Ilgenfritz:2012fw} (see \Tab{tab3}).
Through variation of the fit range we estimate the systematic error 
of the lattice spacing to be smaller than $10\%$. 

To determine the Goldstone pion mass we calculate the corresponding 
correlator, which is given by
\beq
C(n_4)= \sum_{n_1,n_2,n_3} |G(n,0)|^2\,, \label{eq:corr}
\eeq
where $G(n,0)$ is the quark propagator on the lattice. 
We did not apply any smearing in this case.
Though there are in principle benefits by using more complicated sources,
we found that simple point sources are sufficient in our case. 
The effective mass
\beq
M_{eff}(n_4+\frac{1}{2}) = \log\frac{C(n_4)}{C(n_4+1)} \label{eq:eff}
\eeq
was analysed to determine the range where the contribution of higher 
states are negligible, corresponding to a plateau in $M_{eff}(n_4)$.
See middle and right panels of \Fig{fig1}.

In the plateau range we measure the Goldstone pion mass from a fit 
to the correlator (\ref{eq:corr}) of the form
\beq
C_{\pi}(t)= C_0(e^{-Et}+e^{E(t-\Nt)})\,,
\label{eq:pionfit}
\eeq
where $t\equiv n_4$ and $E=m_{\pi}a$.
We obtain a clear plateau in the effective mass and a very good fit
for $t_{min}=7$, as can be seen in \Tab{tab2}.  

Inserting the value of $a$ from (\ref{fm}) in the result for $E$ leads to
\beq
m_{\pi}=175(4)~\mathrm{MeV}, \label{mpi}
\eeq
for $ma=0.0025$ which is, 
as expected from the phenomenological rule $m_\pi^2 \propto m_q$,
about half the value obtained for $ma=0.01$ \cite{Ilgenfritz:2012fw}
(cf. \Tab{tab3}). As we will see below, we now 
have $m_{\pi} \approx T_c(B=0)$.  Therefore, we expect that our
results will be relevant to the physical case of QCD.

\section{Results}
\label{sec:results}

\begin{figure*}[h!t]
\centering
\includegraphics[width=\textwidth]{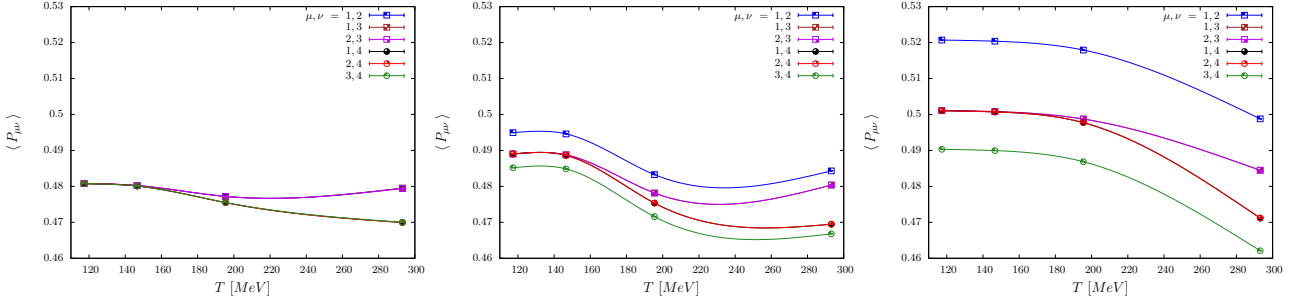}
\caption{Plaquette energies $\langle P_{\mu\nu} \rangle$ vs. 
temperature $T=(a(\beta) \Nt)^{-1}$ 
without magnetic field (left panel), with $qB=0.67~\mathrm{GeV}^{2}$ (middle)
and $qB=1.69~\mathrm{GeV}^{2}$ (right panel) for the different plaquette 
orientations. The lines are only to guide the eye. Computations
were done for $\beta=1.80, am=0.0025, \Ns=32$.}
\label{fig2}
\end{figure*}

We start by discussing the influence of the temperature and 
magnetic field on the different parts $P_{\mu\nu}$ of the gluonic 
action. For convenience we introduce similar variables as 
in \cite{Bali:2013esa}:
\bea
\cE_i^2 & = & \langle P_{4i} \rangle \,, \\
\cB_i^2 & = & \mid \epsilon_{ijk}\mid \langle P_{jk} \rangle \,, \quad j<k\,.
\eea
At $B=T=0$ they are all equal by symmetry. At $B=0, T\neq 0$ 
they fall into two groups, because the fourth direction is not 
equivalent to the other ones:
\beq
\cE_1^2 = 
\cE_2^2 \, = \, 
\cE_3^2\, \le \,
\cB_1^2 = 
\cB_2^2 \, = \, 
\cB_3^2\,.    
\eeq
Introducing a magnetic field in the third direction, for $T\neq 0$ 
the only symmetries left
are rotations in the $(1,2)$-plane. We therefore may define 
\bea
\cE^2_{\parallel} &  \equiv & \cE^2_3\,, \\
\cE^2_{\perp}     &  \equiv & \cE^2_1 =\cE^2_2\,, \\
\cB^2_{\parallel} &  \equiv & \cB^2_3\,, \\
\cB^2_{\perp}     &  \equiv & \cB^2_1 =\cB^2_2\,.
\eea
In \Fig{fig2} we show the results for the four values 
of the temperature $T$, and each of them for the three values of the 
magnetic field $qB$. We give $T$ and $qB$ in physical units via
\Eq{fm}.

We can see the following features from this figure. 
The pattern of the splitting is the same as in our earlier
article \cite{Ilgenfritz:2012fw} and more recently found 
in full QCD \cite{Bali:2013esa},
\beq
 \cB^2_{\parallel} \,\, \geq \,\, 
 \cB^2_{\perp}     \,\, \geq \,\,
 \cE^2_{\perp}     \,\, \geq \,\,
 \cE^2_{\parallel} \,\,.
\eeq
The difference $~\delta_{\perp} \equiv \cB^2_{\perp} -\cE^2_{\perp}$ is 
proportional to a gluonic contribution to the entropy density of the system
(see e.g. \cite{
Engels:1994xj,Boyd:1996bx}). 
Although the latter is not an order parameter it is 
a good indicator for the transition into the deconfinement phase, which
rises the deeper one is penetrating the deconfinement phase. 
In \Fig{fig2} we may compare the $~\delta_{\perp}$-values at fixed $\Nt=6$
(i.e. $T = 195~\mathrm{MeV}~\ageq~T_c(B=0)$) for the three different $qB$-values 
represented in the three panels. We find the relations
$~\delta_{\perp}(qB=0.67~\mathrm{GeV}^2) > \delta_{\perp}(qB=0)
> \delta_{\perp}(qB=1.69~\mathrm{GeV}^2)$. We take this as a first hint
for a non-monotonous behavior of the transition temperature:
$T_c(qB=0.67~\mathrm{GeV}^2) < T_c(B=0) < T_c(qB=1.69~\mathrm{GeV}^2)$.
We shall return to a discussion of plaquette observables 
as a function of $qB$ at the end of this Section.

\begin{figure*}[h!t]
\centering
\includegraphics[width=\textwidth]{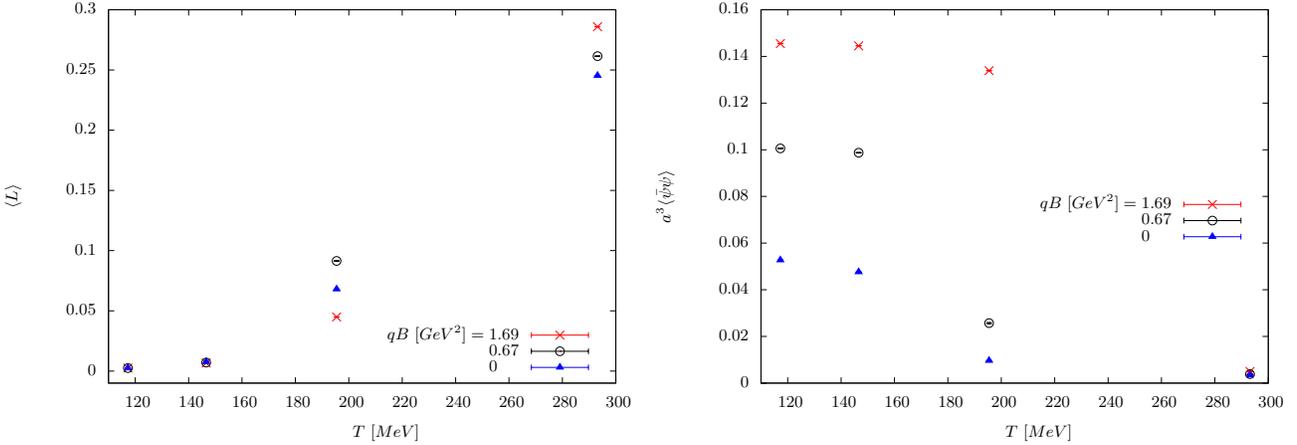}
\caption{Bare Polyakov loop $\langle L \rangle$ (left) and
bare chiral condensate $\langle \bar{\psi}\psi \rangle$ (right)
vs. temperature $T=(a(\beta) \Nt)^{-1}$ 
shown for three values of the magnetic field 
strength at $\beta=1.80,~am=0.0025$ and 
lattice sizes $32^{3} \times \Nt, \; \Nt = 4,6,8,10$.}
\label{fig3}
\end{figure*}
In \Fig{fig3} (left) the expectation value of the unrenormalized Polyakov 
loop $\langle L \rangle$ is shown as a function of the temperature. 
Our values $\Nt=10, 8, 6, 4$ correspond to temperature values $T$, which 
are quite widely spaced. Therefore, we cannot localize the transition e.g. 
for $B=0$ very well. It happens around $T = T_c \simeq 160 - 190$ MeV. 
This means that $T_c \simeq m_{\pi}$, like in QCD. We observe again an 
interesting pattern at $T=195$ MeV ($\Nt=6$), namely that also the Polyakov 
loop does not behave monotonously with the magnetic field (observed 
already in Refs. \cite{D'Elia:2010nq,Ilgenfritz:2012fw}). We will come 
back to that behavior later. We are aware of the fact, that a proper 
renormalization of the Polyakov loop with respect to the $\Nt$-dependence 
will weaken the steep rise with $T$. However, our main conclusions 
concerning the $qB$-dependence at fixed $T$-values will remain unchanged.

In \Fig{fig3} (right) the unrenormalized chiral order parameter 
$a^3 \langle \bar{\psi}\psi \rangle$ is shown versus $T$. 
We observe that for a fixed non-vanishing quark mass it grows monotonously
with the magnetic field at least for the three lower temperature values 
we have investigated. This could be interpreted 
as compatible with an overall {\it magnetic catalysis}. 
However, at $T=195$ MeV, i.e. slightly above $T_c(B=0)$, we detect a 
strong rise of the condensate between $qB=0.67~\mathrm{GeV}^2$ and 
our largest value $1.69~\mathrm{GeV}^2$, while between $qB=0$ and 
$qB=0.67~\mathrm{GeV}^2$ only a small increase of 
$\langle \bar{\psi}\psi \rangle$ is observed.
This behavior shows that the system with rising magnetic field 
strength remains in the chirally symmetric phase until it suddenly
`jumps back' into the chirally broken phase, when the magnetic
field becomes strong enough. This indicates that $T_c(B)$ is rising 
for sufficiently high magnetic field strength. 
At weak magnetic fields, where we saw indications for a lowering of the
critical temperature, the chiral condensate nevertheless does not decrease 
but increases -- although much slower than at lower temperatures 
within the chirally broken phase. 

\begin{figure*}[h!t]
\centering
\includegraphics[width=\textwidth]{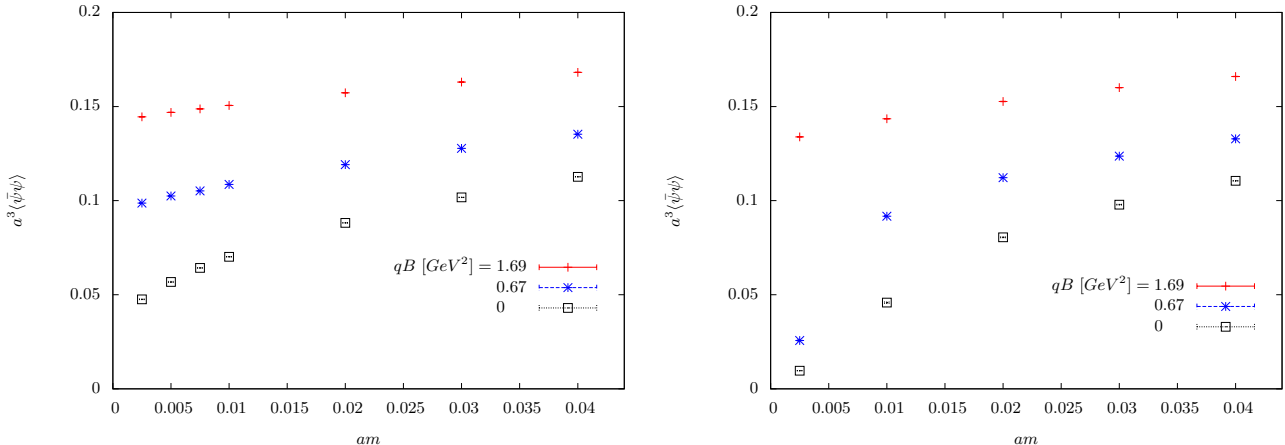}
\caption{Mass dependence of the bare chiral condensate 
$\langle \bar{\psi}\psi \rangle$. Data points are shown 
for $\Nt=8$, i.e. $T=147~\mathrm{MeV}$ (left panel) and 
for $\Nt=6$, i.e. $T=195~\mathrm{MeV}$ (right panel), 
in each case for three values of the magnetic field. 
The simulations were done with $\beta=1.80$ and spatial 
linear lattice extent $\Ns=16$, except for the three 
smallest mass values $am = 0.0025, 0.0050, 0.0075$, where  
$\Ns=32, 24, 20$, respectively, was choosen.}
\label{fig4}
\end{figure*}
We find this pattern confirmed in \Fig{fig4}, where  
$a^3 \langle \bar{\psi}\psi \rangle$ is shown as a function 
of the bare quark mass $ma$ at two temperatures, $T= 147$ and $195$ MeV, 
respectively. One may use these data to extrapolate 
$a^3 \langle \bar{\psi}\psi \rangle$ down to the chiral limit. 
For $T=147$ MeV (left panel) the system is 
clearly in the chirally broken phase for all values of 
$B$ including $B=0$. At the higher temperature $T=195$ MeV 
(right panel) the condensate $a^3 \langle \bar{\psi}\psi \rangle$ 
nicely extrapolates to zero for $qB=0$ corresponding to the chirally
restored phase.
At the intermediate value $qB=0.67~\mathrm{GeV}^2$ there 
seems to be a change of the regime at mass values below $am=0.01$, 
such that the system is consistent with being in the chirally restored 
phase also at this magnetic field strength. On the other hand, 
for the strongest magnetic field strength $1.69~\mathrm{GeV}^2$ 
the data suggest a non-vanishing chiral condensate in the chiral limit. 
Thus, we may conclude that at very strong magnetic field the transition 
temperature grows with $B$, while at fixed $T > T_c(B=0)$ the chiral 
condensate is strongly rising, when the system passes over to the 
chirally broken phase. This is compatible with {\it magnetic catalysis} 
in agreement with various models \cite{Shovkovy:2012zn}. 

\begin{figure*}[h!t]
\centering
\includegraphics[width=\textwidth]{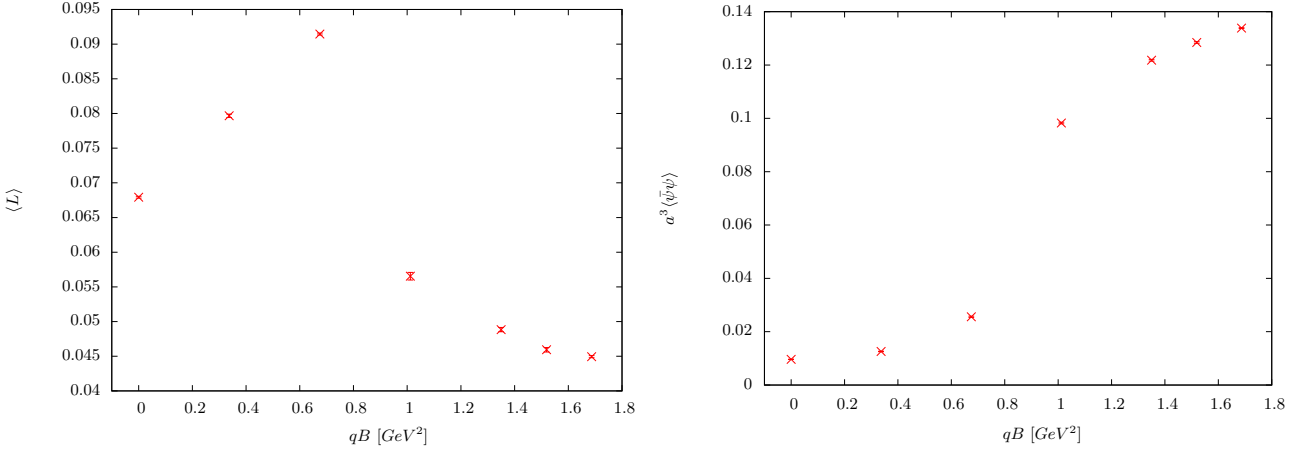}
\caption{Polyakov loop (left panel) and chiral condensate (right panel)
vs. field strength $qB$ at $T=195~\mathrm{MeV}$ 
obtained with $\beta=1.80, am=0.0025$ and lattice size $32^3 \times 6$.}
\label{fig5}
\end{figure*}

In order to study the situation in more detail, we have made simulations at 
fixed $T=195$ MeV ($\Nt=6$) with a few more values of $N_b$. The latter 
correspond to a range of $qB$ between $0$ and $1.69~\mathrm{GeV}^2$. We again
measure the expectation values of the Polyakov loop and the chiral condensate. 
The results are shown in \Fig{fig5}. 
There is a sharp change, which might be related to a
phase transition in the range $0.7\, \mathrm{GeV}^2 < qB < 1.0~\mathrm{GeV}^2$ 
corresponding to $\sqrt{qB}/T \approx 4.5$. This observation
is again supporting a {\it magnetic catalysis} phenomenon.
But for lower magnetic fields we observe a rise of the Polyakov loop with $qB$ 
towards the transition and only then a drop off followed by a monotonous decrease 
at larger field values (compare with our previous comment to \Fig{fig3} (left)).
The rise at low magnetic field values suggests that we are going deeper 
into the deconfinement region, after which the transition brings us back
into the confinement or chirally broken phase. 
The observation of the rise of the Polyakov loop at low magnetic field values 
resembles the pattern discussed in Refs. \cite{Bruckmann:2013oba}, where it
was related to the {\it inverse magnetic catalysis} phenomenon. 

The reader should keep in mind that these data are all obtained at fixed quark 
mass $~am=0.0025$. There the chiral condensate (see the right panel of \Fig{fig5}) 
rises also at a weak field $qB$. However, in this deconfinement range, which 
should be separated from the chirally broken phase by a first order transition,
the chiral condensate is anyway expected to vanish in the chiral limit. Therefore,
the weak monotonous rise of the chiral condensate with $qB$ at the given temperature
$T > T_c(B=0)$ does not mean that an {\it inverse magnetic catalysis} cannot occur
in our model at sufficiently small quark mass. 
 
\begin{figure*}[h!t]
\centering
\includegraphics[width=\textwidth]{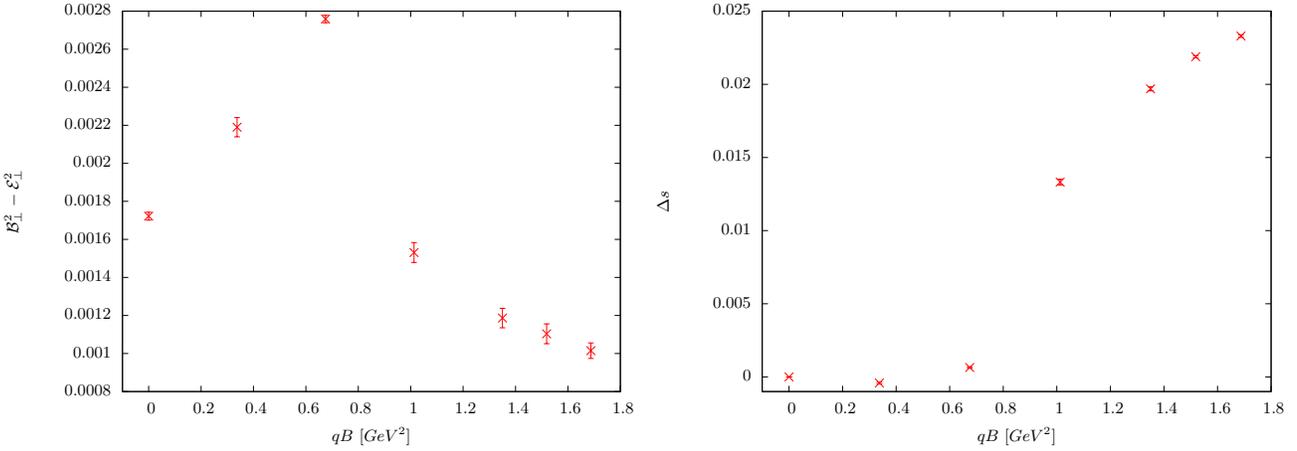}
\caption{The difference of plaquette energies 
$~\delta_{\perp}=\cB^2_{\perp} -\cE^2_{\perp}$ 
(left panel) as well as the purely gluonic contribution $\Delta s$ 
to the interaction measure (right panel) vs. field strength $~qB~$ at 
$~T=195~\mathrm{MeV}$ as in \Fig{fig5}.}
\label{fig6}
\end{figure*}
In order to gain more information let us come back to the plaquette
variables. In \Fig{fig6} (left) we show the plaquette energy difference 
$~\delta_{\perp} = \cB^2_{\perp} -\cE^2_{\perp}$
as a function of $qB$ for the same temperature
as in \Fig{fig5}. Where we saw a rise of the Polyakov loop with $qB$, 
we observe now also a rise of the difference $~\delta_{\perp}$, 
indicating again that we are going deeper into the 
deconfinement phase. Contrary to that, $~\delta_{\perp}$
decreases at larger $qB$-values, where we are
driven by the magnetic field into the confinement phase.  
Another observable derived from the average plaquette 
variables is the gluonic contribution to the subtracted 
{\it interaction measure} defined as
\beq
\Delta s \equiv \langle~ P ~\rangle(T,B) - \langle~ P ~\rangle(T,B=0)\,,
\eeq  
where $\langle~ P ~\rangle$ includes the average over all plaquette
orientations. We do not consider a factor given by the derivative of the
$\beta$-function with respect to the lattice spacing, because it is 
irrelevant in our fixed-scale method. We have plotted $\Delta s$ versus
$qB$ in the right part of \Fig{fig6}. At low values of $qB$ we observe
a small decrease, whereas at larger magnetic field strength $\Delta s$ 
is rising. The decrease at low $qB$ seems also to be compatible with a similar 
observation discussed in \cite{Bali:2013esa}, where it was related to 
{\it inverse magnetic catalysis} .   

\begin{figure}[h!t]
\centering
\includegraphics[width=.45\textwidth]{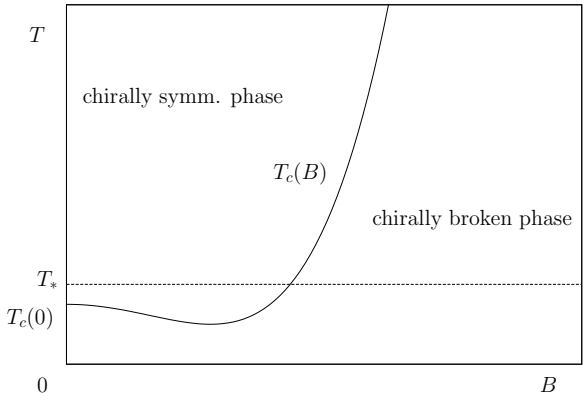}
\caption{Conjectured B-T phase diagram at fixed mass $am=0.0025$. 
The horizontal line $T=T_*=\mathrm{const.}$ indicates the path of 
simulations at $T=195~\mathrm{MeV}$ as in Figs. \ref{fig5} and 
\ref{fig6}.}
\label{fig7}
\end{figure}
Our observations above seem to indicate a decrease of $T_c$ with rising
but small $qB$. At large $qB$ the transition temperature $T_c$ definitely 
rises as expected in the case of a magnetic catalysis. 
In \Fig{fig7} we conjecture a $B-T$ phase 
diagram, which might clarify the situation. In order to prove it, 
further simulations at somewhat smaller temperatures and/or smaller 
quark mass would be helpful. If it proves to be true then -- for the 
same $am=0.0025$ or even lower mass -- one should find a path 
$~T=\mathrm{const.}~<~T_c(B=0)~$ for which at $qB=0$ the system is in 
the confinement (chirally broken) phase. With increasing $qB$ one passes 
then the chirally restored phase, i.e. the deconfinement or chiral transition 
twice, and ends up again in the confinement phase. Along such a
path in the phase diagram the chiral condensate should  
decrease with $qB$ when entering the chirally restored phase.   
This would mean the existence of {\it inverse magnetic catalysis} 
also in two-color QCD.  

Let us finally notice that in the recent 
papers \cite{Andersen:2012jf,Orlovsky:2013xxa}
similar scenarios as proposed here were obtained for the cases 
$N_f=2$ and $N_f=2+1$, which differ from ours by a smooth crossover behavior. 

\vspace*{0.5cm}

\section{Conclusions}
\label{sec:conclusions}

In this article we have described an investigation of two-color QCD 
at finite temperature in a constant external magnetic field. 
We spezialized to the case of four flavors of equal charge $q$ 
implementing staggered fermions on the lattice without employing the 
fourth root trick.
We have performed lattice simulations using a fixed-scale approach 
so that we do not need to know the beta-function and the
dependence of the renormalization constants on the bare coupling constant. 
The simulations were carried out at a lattice spacing $a \approx 1/(6~T_c(0))$,
where $T_c(0)$ is the critical temperature of the finite temperature transition
at vanishing magnetic field. Furthermore, we used a fixed bare quark mass 
which is four times smaller than in our previous work \cite{Ilgenfritz:2012fw}.
This means that now the Goldstone meson mass is $m_{\pi} \approx T_c(0)$
similar to the physical case of QCD. We have also taken some data at 
larger quark masses to be able to extrapolate to the chiral limit.

We find that at sufficiently large magnetic fields there is 
{\it{magnetic catalysis}}, i.e. the chiral order parameter and the critical 
temperature are increasing with increasing magnetic field strength. 
This is in agreement with predictions by many models. The result is, 
however, apparently different from that of Ref. \cite{Bali:2011qj} 
close to the physical point of QCD, where one finds
 {\it inverse magnetic catalysis} in the crossover region, i.e. the chiral
order parameter is not increasing monotonically with the magnetic field 
strength, and as a consequence the transition temperature decreases.

In our case of two-color QCD with $N_f=4$ dynamical fermion degrees
of freedom a real phase transition is expected in contrast to a
smooth crossover, and one may therefore expect that the deconfinement 
and chiral transition should coincide. Therefore,
a priori an inverse magnetic catalysis phenomenon could be absent.
However, as we showed, there are indications that for
weak magnetic field the critical temperature $T_c(B)$ is decreasing
with rising $B$. If so, for fixed sufficiently small quark mass and fixed 
temperature $T~\aleq~T_c(B=0)$ we should be able to find a trajectory in the 
phase diagram along which one passes from the chirally broken phase 
with its large chiral condensate through the chirally restored phase 
with a suppressed chiral condensate again into the chirally broken 
phase at larger $qB$-values. In this way we should observe a real 
inverse magnetic catalysis replacing the crossover behavior 
observed in QCD close to the physical point. 

Since our theory is different from QCD the behavior does not have to be the 
same, but the response of the system to a strong magnetic field should be 
to some extent model independent for theories with similar chiral properties.
We note, however, that the magnetic field strength at which we see a clear
signal of magnetic catalysis is (to the extent one can compare scales in 
different theories) larger than that investigated in \cite{Bali:2011qj}. 
At a magnetic field strength similar to those used in \cite{Bali:2011qj} 
our data are consistent with the possibility of an inverse magnetic 
catalysis scenario in the sense described above.

In \cite{Bruckmann:2013oba} e.g. it is claimed that the inverse 
magnetic catalysis is due to the coupling of the magnetic field to the 
sea quarks. This also gives rise to an increase in the Polyakov loop 
with the magnetic field strength, which is an effect that we see in our
calculations up to a critical magnetic field. Going beyond the latter 
the Polyakov loop suddenly drops to a value near zero, and the system 
enters the confined phase. This critical magnetic field is stronger 
than that used in \cite{Bali:2011qj,Bruckmann:2013oba}.

It would be very interesting to have results from QCD calculations 
at stronger magnetic fields to see if the phase diagram of \cite{Bali:2011qj} 
extends to the one we propose in \Fig{fig7}, or if the inverse magnetic 
catalysis persists for all values of the magnetic field strength.

A further simulation of our model at a somewhat lower temperature
would be helpful to pinpoint the critical line in the phase diagram. 
Investigations at different quark masses $ma$ would be important, 
because the phase transition line is expected to depend on the quark mass.  
Our calculations should be also extended to smaller scales $a$ 
in order to extrapolate to the continuum limit.

\vspace*{0.5cm} 
\section*{Acknowledgments}
Useful discussions with M. D'Elia, T. Kovacs, E. Laermann and
a correspondence with L. McLerran are gratefully acknowledged. We thank
J.O. Andersen and V. Orlovsky for bringing their papers 
\cite{Andersen:2012jf} and \cite{Orlovsky:2013xxa}, respectively, to our
attention. We express our gratitude to F. Bruckmann and G. Endrodi 
for a critical reading of our manuscript and for useful comments, 
and F. Burger for continuous technical help and advice, in particular 
for running our CUDA codes on a PC cluster with GPU's.  

\bibliographystyle{apsrev}

\end{document}